 \documentclass{agujournal}

\journalname{Geophysical Research Letters}
\usepackage{hyperref}
\usepackage{comment}
\usepackage{siunitx}

\begin{document}

%
%


\title{Representing ill-known parts of a numerical model using a machine learning approach}

%
%




\authors{Julien Brajard\affil{1,2}, 
Anastase Charantonis\affil{3,4}, 
J\'er\^ome Sirven\affil{1}}


\affiliation{1}{Sorbonne University, CNRS-IRD-MNHN, LOCEAN, Paris, France}
\affiliation{2}{Nansen Environmental and Remote Sensing Center (NERSC), Bergen, Norway}
\affiliation{3}{\'Ecole Nationale Sup\'erieure d'Informatique pour l'Industrie et l'Entreprise, \'Evry, France}
\affiliation{4}{Laboratoire de Math\'ematiques et Mod\'elisation d'\'Evry, \'Evry, France}




\correspondingauthor{Julien Brajard}{julien.brajard@locean-ipsl.upmc.fr}




\begin{keypoints}

\item We present a methodology combining a physical-driven and a data-driven model into a hybrid model applied to a shallow-water model.
\item The forcing, dissipative and diffusive terms are represented in this hybrid model by a neural network.
\item The hybrid model is able to reproduce the long term conservation properties of the target simulation and its mean state, as well as accurately predict its short term evolution.
\end{keypoints}

%
%


\begin{abstract}
In numerical modeling of the Earth System, many processes remain unknown or ill represented 
(let us quote sub-grid processes, the dependence to unknown latent variables or the non-inclusion of complex dynamics in numerical models) but sometimes can be observed.
This paper proposes a methodology to produce a hybrid model combining a physical-based model (forecasting the well-known processes) with a neural-net model trained from observations (forecasting the remaining processes).
The approach is applied to a shallow-water model in which the forcing, dissipative and diffusive terms are assumed to be unknown.
We show that the hybrid model is able to reproduce with great accuracy the unknown terms (correlation close to 1). For long term simulations it reproduces with no significant difference the mean state, the kinetic energy, the potential energy and the potential vorticity of the system. Lastly it is able to function with new forcings that were not encountered during the training phase of the neural network.

\end{abstract}

%
%

%


%
%
%
%

\section{Introduction}
The temporal evolution of the atmosphere or ocean can be represented by a set of equations based on the fundamental laws of physics and empirical relations deduced from observations. These equations are complex and numerical methods have been developed to solve them. The quality and complexity of these numerical models has iteratively increased, culminating in the actual ocean or atmosphere general circulation models~\citep{Eyring2016OverviewOrganization}. In oceanography, the most complex models currently use a grid with an horizontal resolution of about 10 km, (e.g.~\citet{Madec2015NEMOEngine}) with 50 vertical levels on a square basin with sides of 5000 km, corresponding to $1.25\times 10^7$ grid points. When the model is run to simulate a year, with 1 hour time steps, the information brought by the numerical experiment, for a single variable, is contained in a vector of around $\sim10^{11}$ values.

This numerical approach can be described using formal mathematical notations in the following way. A vector $\mathbf{x} \in \mathbb{R}^n$ (typically $n \sim 10^{7-8}$) defines the discretized state of the system. A dynamical model is a system of differential equations which can be solved to compute vectors $\mathbf{x}_i \in \mathbb{R}^n$ for time index $ i \in {1, \dots , p } $ using the information already known: $ \mathbf{x}_i = \Phi (\mathbf{x}_{i-1}, \ldots , \mathbf{x}_{0} ) $ where $\Phi $ is a highly complex function obtained by discretizing the basic equations of the model. If $\Phi$ is derived from physical principles, the model is said to be "physically driven". The miss-representation  of sub-scale phenomena and the imperfect resolution of some non-linear processes is an inherent part of the numerical approach which always has to make a trade-off between accuracy and computation cost. To remedy these deficiencies, the ill-resolved parts of the models are represented through parameterization schemes and simplified or empirical models~\citep{Randall2007ClimateEvaluation}.

In this paper we use machine learning and a data-driven approach to build a hybrid model, which combines physically and data driven terms, in order to reproduce the processes that are inaccurately represented in the original model. Several works, which used data to derive models based on machine learning and deep learning algorithms, have already been conducted. Approaches aiming at emulating an entire model given a set of perfect (without noise) observations ~\citep{Pathak2017UsingData,Pathak2018Model-FreeApproach,Fablet2017} were applied on chaotic low-dimension dynamical systems (such as the 3 variable Lorenz model~\citep{Lorenz1963DeterministicFlow}, the 40 variable Lorenz model~\citep{Lorenz1998OptimalModel} and the Kuramoto-Sivashinsky model~\citep{Kuramoto1976PersistentEquilibrium}). "Physical-oriented" deep-learning architectures have also been proposed for forecasting or nowcasting with no explicit numerical models~\citep{debezenac2017,Shi2015ConvolutionalNowcasting}. For perfect data, it has been proposed that machine learning approaches can help estimating unknown parameters or unknown parameterizations of a numerical model~\citep{Schneider2017a}. Deep-learning was also applied to infer sub-grid parameterizations ~\citep{Bolton2019ApplicationsParameterisation, Rasp2018DeepModels.}.

Our approach aims at representing the ill-known part of a numerical model. The main contributions of this paper are to address the case of noisy data, to aim at conducting long-term simulations that conserves the properties of the underlying dynamics and to evaluate the generalization skill of the hybrid model. We show that it is necessary to add physical constraints to the hybrid model in order to achieve these objectives.

In the machine learning approach, we rewrite the numerical model 
\begin{linenomath*}
\begin{equation}
\mathbf{x}_i = \Psi (\mathbf{x}_{i-1}, \ldots , \mathbf{x}_{0} , \boldsymbol{\theta}(i) )
\label{eq:model}
\end{equation}
\end{linenomath*}
where $\boldsymbol{\theta} (i) $ represents the the ill-known part of the  numerical model at iteration $i$. It can be estimated using an empirical model; for this we suppose the estimator $\hat{\boldsymbol{\theta}} $ of $\boldsymbol{\theta}$ depends on the state of system: 
\begin{linenomath*}
\begin{equation}
\hat{\boldsymbol{\theta}}(i) = g (\mathbf{x}_{i-1}, \ldots , \mathbf{x}_{0} ) 
\label{eq:empirical}
\end{equation}
\end{linenomath*}
where $g$ is a unknown function that the machine learning algorithm must determine by means of non-linear regressions in a high dimensional space. In the following we will consider the particular case in which $g$ depends only on the previous state: $\hat{\boldsymbol{\theta}}(i) = g (\mathbf{x}_{i-1})$ The function $g$ and thus $\hat{\boldsymbol{\theta}}$ can be estimated thanks to external data coming from two sources:
\begin{itemize}
\item Direct or indirect observations of the system (e.g. in-situ observations or satellite data)
\item High-resolution simulations (they can be carried out to resolve the ill-resolved processes on coarser resolutions).
\end{itemize}
Note that the access of these data are limited in space and time. In this paper we firstly use external data (see section 2.2) to calibrate the $g$ function and secondly integrate the neural-network model in absence of these external data.

In the last years,  big data and machine learning techniques (deep-learning, neural networks, ...) have shown impressive skills in forecasting and classifying complex behaviors~\citep{lecun2015}. In this work, we use a convolutional neural network~\citep{Goodfellow-2016} to compute the $g$ function (see Eq.~\ref{eq:theta}). We try to evaluate the capacity of the neural-network model to produce long-term simulations similar to those produced by a medium-complexity shallow-water model (described in section \ref{sec:swm}). The computations are constrained by weakly noised observations of model outputs.

The paper is organized as follows. Section 2 presents the physical model and section 3 the method. The last section shows the results based on synthetic data generated with or without noise, with two different wind forcings.

\section{The model and the reference experiment}
\label{sec:swm}
\subsection{The shallow-water model}
We consider a one and a half layer shallow-water model in the $\beta $ plane. Noting $x$ and $y$ the zonal and meridional coordinates we have
\begin{linenomath*}
\begin{eqnarray}
\label{eq:swmodel}
\partial_tu & = &
+ (f + \zeta ).v  - \partial_x (\frac{u^2+v^2}{2} + g^*.h) + \boldsymbol{\theta}_u\nonumber \\
\partial_tv & = &
-  (f + \zeta ).u  - \partial_y (\frac{u^2+v^2}{2} + g^*.h) +
                  \boldsymbol{\theta}_v \\
\partial_th & = &
- \partial_x(u(H+h)) - \partial_y(v(H+h)) \nonumber
\end{eqnarray}
\end{linenomath*}
The vector $\boldsymbol{\theta} =(\boldsymbol{\theta}_u,\boldsymbol{\theta}_v)$ contains the dissipation, the diffusion, and the forcing and is assumed to be unknown; $u$,$v$ are the zonal and meridional velocities, $h$ the thickness anomaly of the active layer, $f$ the Coriolis parameter ($f = f_0 +\beta y $, $f_0=$ \SI{.5e-5}{\per \second}, $\beta$=\SI{2.11e-11}{\per \meter \per \second}), $\zeta = \partial_y v - \partial_x u $  is the vorticity, $g^*=$\SI{0.02}{\meter \per \second \squared} is the reduced gravity, and $H=$\SI{500}{\meter} is the mean thickness of the active layer. The domain size is \SI{1600}{\kilo\meter} $\times$ \SI{1600}{\kilo\meter} with a grid point of \SI{20}{\kilo\meter}.

The equations are discretized on a C-grid~\citep{Arakawa1981AEquations} with a resolution of 20 km, largely inferior to the mean Rossby deformation radius of the model, around 620 km. A leap-frog scheme for temporal discretization is used, combined with an Asselin
filter~\citep{Asselin2007FrequencyIntegrations} with a time step of \SI{1800}{\second}. This produces a recurrence relation similar to the  one summarized in equation (1).

\subsection{Reference experiment}
In order to train our neural network (see Section~\ref{sec:nn}) to determine the function $g$ and thus to reproduce $\boldsymbol{\theta}$, we analytically define a simulated "truth":
\begin{linenomath*}
\begin{eqnarray}
\boldsymbol{\theta}_u^{\rm target}& =  g^{\rm target}_u(u,h,\tau_x)  = & \frac{\tau_x}{\rho_0(H+h)} - \gamma . u + \nu\Delta u \nonumber\\
\boldsymbol{\theta}_v^{\rm target} &=  g^{\rm target}_v(v,h,\tau_y)  = & \frac{\tau_y}{\rho_0(H+h)} -
                       \gamma . v + \nu\Delta v 
\label{eq:theta}
\end{eqnarray}
\end{linenomath*}
where $\Delta $ is the Laplace operator, $\rho_0=$\SI{1000}{\kg\per \cubic\meter}, $\gamma=$\SI{2e-7}{\per \second}, $\nu=$\SI{0.72}{\meter\squared\per\second}. The meridional wind stress $\tau_y $ is null and the zonal wind stress $\tau_x$ is defined by :
\begin{linenomath*}
\begin{equation}
\tau_x(y) = \tau_0 \sin(2\pi (y-y_c)/L_y)
\label{eq:wind}
\end{equation}
\end{linenomath*}
where:
$\tau_0$ is in the standard case equal to \SI{0.15}{\newton \per \meter\squared} ($y_c = 0$ km and $L_y = $\SI{1600}{\km}).

The fully specified model (Eq.~\ref{eq:swmodel} and Eq.~\ref{eq:theta}) is the same as in~\citet{Krysta2011AOcean} with changes in parameter values to ensure the presence of typical dynamical structures as eddies. No slip boundary conditions are applied.

Note that the unknown terms $\boldsymbol{\theta}$ in our model are diverse; each has a different impact on the evolution of the system's state. On the one hand the wind forcing varies in space and structures the mean state of the shallow-water model. On the other hand, the dissipation and diffusion depend on the state variable and dissipate energy. A quasi-stationary state of the system is obtained because of a balance between these terms after a spin-up period. Addressing the problem of the capacity of a neural network to represent these heterogeneous processes in a model is one of the main objectives of this work.

The observations $\boldsymbol{\theta}^{\rm target}$ used to constraint the neural network are generated by the fully-specified mode. In a real test case, we should rely on external data as discussed in the introduction.

\section{Methods}
\subsection{Neural Network}
\label{sec:nn}
A convolutional neural network (CNN) is used; it has shown abilities to represent high-level patterns in a image~\citep{Goodfellow-2016}. In our case, this skill helps us to represent differential operators (e.g. Laplace operator) and non-linear functions. It is constituted with $L_1, \cdots , L_d$ successive layers; each layer $L_m$ computes $p_m$ convolutions:
\begin{linenomath*}
\begin{equation}
z_{ijk}^m = s \left( 
	\sum_{k'=1}^{p_{m-1}}
	\sum_{(i'j') \in \mathcal{N}^m(ij)}
    W^m_{i'-i,j'-j,k}.z_{i'j'k'}
    \right)
\end{equation}
\end{linenomath*}
where:
\begin{description}
\item [$(ij)$] is the index of a grid point in the domain
\item [$k$] is the index of the convolution performed by layer $L_m$ ($k \in [1..p_m]$).
\item [$z_{ijk}^m$] is the computed result of the convolution of the layer $L_m$
\item [$\mathcal{N}^m(ij)$] is a set of index points in the vicinity of the point $(ij)$. It corresponds the size of the convolution kernel (typically $3\times 3$)
\item [$W^m_{i'-i,j'-j,k}$] is the values (or so-called \emph{weights}) of the convolution kernel.
\item $s$ is a non-linear function that introduces non-linearities in the computation. Except for the last layer where $s$ is the identity function $s(z)=z$, the chosen function is the so-called rectifier function : $s(z) = max(0,z)$
\end{description}
The last layer (the output layer) gives an estimation  of $ \hat{\boldsymbol{\theta}}_u $  and $ \hat{\boldsymbol{\theta}}_v $. 
Each layer $L_m$ takes inputs from the preceding layer $L_{m-1}$, except for the first layer that takes as input the state variables of the model $\mathbf{x} \in \{(u,h,\tau_x),(v,h,\tau_y)\}$ (see Eq.~\ref{eq:theta}).

A CNN can thus be represented as a function $\hat{\boldsymbol{\theta}}=g^\mathit{nn}_\theta(\mathbf{x},\mathbf{W})$ where $\mathbf{W}$ represents the weights of all the convolution kernels. The learning phase of the neural networks consists in adjusting $\mathbf{W}$ in order to minimize the discrepancy between $\hat{\boldsymbol{\theta}}$ and $\boldsymbol{\theta}^{\rm target} \in \{\boldsymbol{\theta}_u^{\rm target},\boldsymbol{\theta}_v^{\rm target}\}$(see Eq.~\ref{eq:theta}). The minimization is performed iteratively given a training dataset of matching examples ($\mathbf{x}$,$\boldsymbol{\theta}^{\rm target}$).

The selection of the architecture (number, size of layers, size of the kernels, activation functions, ...) of our CNN is the result of a cross-validation process. Its parameters are summarized in Table~\ref{tab:nn}. Note that the first layer of the neural network uses kernel of size $3 \times 3$. It makes our model able to mimic first-order finite differences. Had the scheme we were trying to approximate corresponded to a higher order of finite differences, we should have used a larger convolution filter size, or have added extra convolution layers. The last layer is equivalent to a dense connected layer with shared weights at each grid point. In our case, we have used the same CNN architecture for estimating $\boldsymbol{\theta}_u$ and $\boldsymbol{\theta}_v$. 

 \begin{table}
 \caption{\label{tab:nn}Architecture of the neural-net $g^\mathit{nn}_\theta$}
 \centering
 \begin{tabular}{r c}
 \hline
  Input Size  & $80 \times 80 \times 3$ \\
Number of  layers & 2 \\
Number of units in each layer & 32, 1 \\
Size of the kernel in each layer & $(3\times3)$,$(1\times 1)$\\
Activation function in each layer & ReLU, linear \\
Output size  & $80 \times 80 \times 1$ \\
Total number of weights & 929 \\
\hline
 \end{tabular}
 \end{table}

\subsection{Datasets}
Several datasets were generated using the fully-specified shallow-water numerical model in order to train the neural networks and test their generalization capacity. These datasets are summarized in Tab.~\ref{tab:sets}. In each dataset, there are two subsets: (i) the $u$ subset representing matches between $\mathbf{x}_u=(u,h,\tau_x)$ and $\boldsymbol{\theta}_u$ and (ii) the $v$ subset representing matches between $\mathbf{x}_v=(v,h,\tau_y)$ and $\boldsymbol{\theta}_v$.

Two training sets are used for training. Dataset {\texttt{train\_nonoise}} represents the ideal case in which the knowledge 
of the matching between $\mathbf{x}$ and $\boldsymbol{\theta}^{\rm target}$ is known with a perfect accuracy. The examples are 480 complete fields of ($\mathbf{x}$,$\boldsymbol{\theta}^{\rm target}$) computed from Eq.~\ref{eq:theta} during a 40 years simulation by extracting snapshots at a frequency of approximately 1 month. Dataset {\texttt{train\_noise01}} represents a case in which the matching is imperfectly known. The snapshots are the same as for {\texttt{train\_nonoise}} with an added noise for each parameter of the dataset:
\begin{linenomath*}
\begin{equation}
x_{noise} = x + \epsilon \times \sigma_x
\label{eq:noise}
\end{equation}
\end{linenomath*}
where $x$ stands for a variable in the dataset $\in \{u,v,h,\tau_x,\tau_y,\boldsymbol{\theta}^{\rm target}_u,\boldsymbol{\theta}^{\rm target}_v\}$, $\epsilon$ is a random value drawn for a Normal distribution of mean 0 and standard deviation 0.1 and $\sigma_x$ is the standard deviation in the dataset of the corresponding variable $x$.

We aim at testing the skill of the CNN compared to a fully-known numerical model. Therefore the test datasets do not contain noise. The first dataset {\texttt{test\_windstd}} is produced exactly in the same conditions as {\texttt{train\_nonoise}} except that the initial conditions of the simulation are taken to be the final state of the simulation used to generate {\texttt{train\_nonoise}}. It ensures that the model states used for testing are different than those used for training.

In order to test the generalization capacity of the hybrid model, a second dataset {\texttt{test\_windlow}} was produced. It was obtained by changing the wind forcing compared to the first dataset: the wind intensity in Eq.\ref{eq:wind} is set to $\tau_0=0.10 m.s^{-2}$. In this case, the hybrid model is forced by a wind forcing that was never encountered during the training. Note that, among the parameters, $\tau_0$  has the strongest impact on the system. 

 \begin{table}
 \caption{\label{tab:sets}Summary of the datasets used for training and testing the neural networks.}
 \centering
 \begin{tabular}{r c c }
 \hline
  Name  & Size$^{a}$ & Notes  \\
 \hline
   \texttt{train\_nonoise}  & 480  &  Eq.~\ref{eq:theta}\\
   \texttt{train\_noise01}  & 480  &  Eq.~\ref{eq:noise} \\
   \texttt{test\_windstd} & 120 & Eq.~\ref{eq:theta} \\  
   \texttt{test\_windlow} & 120 & Eq.~\ref{eq:wind} with $\tau_0=0.10$  \\
 \hline
 \multicolumn{3}{l}{$^{a}$Number of snapshots contained in the dataset}
 \end{tabular}
 \end{table}

\subsection{\label{sec:diag}Validation diagnostics}
Due to the large variability and the chaotic nature of the model, we cannot compare a reference simulation with the neural-network model on a term by term basis. To validate a model run, then, some mean conservative quantities that represent the physical characteristics of the system being modeled are computed : 
the Kinetic Energy (KE), the Potential Energy (PE) and the potential vorticity (PV):
\begin{linenomath*}
\begin{eqnarray}
KE& = & \frac{1}{2S} \int_S (H+h)(u^2+v^2) ds \nonumber\\ 
PE& = & \frac{1}{2S} g^*  \int_S (H+h)^2 ds \\
PV& = & \frac{1}{S} \int_S \frac{f + \zeta }{H+h} ds \nonumber \label{eq:conserv}
\end{eqnarray}
\end{linenomath*}
where $S$ is the surface of the domain. The evolution of these four quantities in the fully specified model (using Eq.~\ref{eq:theta}) is compared with their evolution in the hybrid model.

\section{Results}
\subsection{Training and testing the neural networks}
For each target parameter $\boldsymbol{\theta}_u$, $\boldsymbol{\theta}_v$, one training was performed for each of the training datasets described in Table~\ref{tab:sets}. Thus, a total of 4 neural networks were trained. The performances, calculated on the test datasets, are shown in Table~\ref{tab:apriori}.

For each training, the {\it a priori} performances show that the neural network is able to reproduce $\boldsymbol{\theta}_u$ and $\boldsymbol{\theta}_v$. Note that $\boldsymbol{\theta}_v$ is easier to reproduce than $\boldsymbol{\theta}_u $ because the meridional component of the wind stress is null. This leads to a significantly lower RMSE for $\boldsymbol{\theta}_v$ than for $\boldsymbol{\theta}_u$. Neither the addition of noise in the training set, nor the test with a different wind intensity significantly degrade the accuracy of the CNN.

\begin{table}
 \caption{\label{tab:apriori}{\it A-priori} performance of the neural networks on test-dataset}
 \centering
 \begin{tabular}{r c c c c}
 \hline
  Train set  & Test stet$^{a}$  & parameter$^{b}$ & RMSE$^{c}$ & corr.$^{d}$ \\
 \hline
   \texttt{train\_nonoise}  & \texttt{test\_windstd}  & $\boldsymbol{\theta}_u$ & $1.07\cdot10^{-9}$  & 1.000  \\
   \texttt{train\_nonoise}  & \texttt{test\_windstd}  & $\boldsymbol{\theta}_v$ & $5.24\cdot10^{-11}$ & 1.000  \\
\texttt{train\_noise01}  & \texttt{test\_windstd}  & $\boldsymbol{\theta}_u$ & $3.27\cdot10^{-9}$  & 0.9999 \\
\texttt{train\_noise01}  & \texttt{test\_windstd} & $\boldsymbol{\theta}_v$ & $2.04\cdot10^{-9}$  & 0.9992 \\
   \texttt{train\_nonoise}  & \texttt{test\_windlow}  & $\boldsymbol{\theta}_u$ & $2.68\cdot10^{-9}$  & 0.9998 \\
   \texttt{train\_nonoise}  & \texttt{test\_windlow}  & $\boldsymbol{\theta}_v$ & $4.71\cdot10^{-11}$ & 1.000  \\
   \texttt{train\_noise01}  & \texttt{test\_windlow}  & $\boldsymbol{\theta}_u$ & $3.83\cdot10^{-9}$  & 0.9997 \\
   \texttt{train\_noise01}  & \texttt{test\_windlow}  & $\boldsymbol{\theta}_v$ & $1.78\cdot10^{-9}$  & 0.9987 \\
 \hline
 \multicolumn{5}{l}{$^{a}$ Training set used for the neural network}\\
 \multicolumn{5}{l}{$^{b}$ predicted parameter}\\
  \multicolumn{5}{l}{$^{c}$ root-mean-square-error}\\
    \multicolumn{5}{l}{$^{d}$ correlation coefficient}
 \end{tabular}
 \end{table}

\subsection{Simulation using the hybrid model}
Using the neural network trained with dataset {\texttt{train\_noise01}}, two 10 year simulations with different wind forcing ($\tau_0$ = 0.1 and $\tau_0$ = 0.15 in Eq.~\ref{eq:wind}) were made with the hybrid model (Eq.~\ref{eq:model}). We recall that the forcing ($\tau_0$ = 0.1) was never encountered during the training phase of the CNN; thus it allows us to test the generalization skill of the hybrid model. 

As a comparison, the  fully-specified numerical model was also run using the analytical expression in Eq.~\ref{eq:theta} which is considered as our "true" reference simulation.

\begin{figure}[ht]
\centering
\includegraphics[width=20pc]{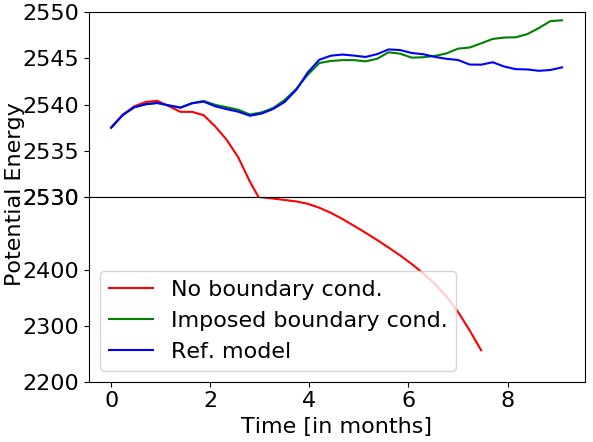}
\caption{Evolution in time of the Potential Energy for the reference model (in blue), the hybrid model with no boundary constraints (in red) and the hybrid model with imposed strictly null velocity orthogonal to the border. Note that the scale of the y-axis is not regular in order to visualize the different orders of magnitude. The unconstrained simulation after 8 months is not shown because NaN (Not A Number) values were encountered.}
 \label{fig:diverge}
\end{figure}

To ensure that the simulation does not diverge, it is essential to impose the boundary conditions to the data-driven terms $\boldsymbol{\theta}$ in the hybrid model. This essential point is illustrated in Fig.~\ref{fig:diverge}. In one case (red curve), the hybrid model is used in the whole domain (including the boundaries) to predict $u$ and $v$. In the other case, the velocity orthogonal to the border of the domain is set to zero. Even though the unconstrained neural-net is giving a very small error for boundary velocity (root mean square error of $\sim10^{-9}$), this error suffices in damaging the mass conservation and makes the CNN diverge after 2-3 months of simulation. Fig.~\ref{fig:diverge} also shows that the hybrid model presents forecast skills (up to 1 month for the unconstrained model, and up to 5-6 months for the constrained model). In the following, only the constrained model was used.

\begin{figure}[ht]
\centering
\includegraphics[width=20pc]{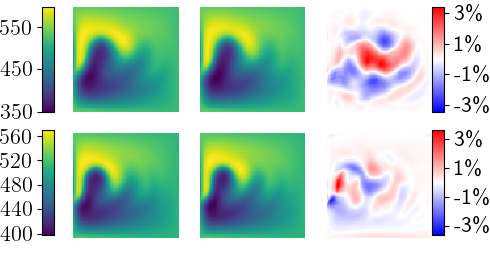}
\caption{Mean state of $H+h$ for the hybrid model simulation (left panel), reference simulation (central panel) and the relative difference between both (right panel). The upper row is the simulation performed with $\tau_0$=0.15 (used also during the training) and the lower row is the simulation performed with $\tau_0=0.10$}
 \label{fig:mean_state}
\end{figure}

For different experiments, the thickness $H+h$ averaged over 10 years is shown in Figure~\ref{fig:mean_state}. The  conservative parameters defined in Eq.~\ref{eq:conserv} were computed every 5 months of the 10 year simulation then their time average (with confidence interval at 99\%) are shown in Table~\ref{tab:diags}. First, the two simulations with different wind forcing exhibit significant differences for all the conservative parameters (with a 99\% confidence level) and the mean state. In comparison, differences between the reference simulation and the hybrid model simulation are not significant (with a 99\% confidence level). Therefore, all the properties considered to characterize the simulation are well reproduced by the neural networks simulations.

Even though the hybrid model was forced with a wind intensity that was never "seen" during the training phase, the diagnostic properties are fairly reproduced. This shows that the neural network model has generalization skills. 

\begin{table}
 \caption{\label{tab:diags}Long-term diagnostics of the simulations for the reference model (ref) and the hybrid model (hybrid). For the definitions of the parameters, see Eq.~\ref{eq:conserv}. The confidence intervals are computed with a confidence level of 99\%}
 \centering
 \begin{tabular}{|r| c c| c c|}
 \hline
  & \multicolumn{2}{c|}{$\tau_0 = 0.15$} &  \multicolumn{2}{c|}{$\tau_0 = 0.1$} \\
\cline{2-5}
Parameter & ref & hybrid & ref & hybrid\\
 \hline
 KE & $24.969 \pm 1.06 $ & $24.997 \pm 1.06$ 
 & $11.80 \pm 0.32$ & $11.99 \pm 0.36$\\
 PV  $\times 10^{-8} $& $7.33 \pm 0.02$ & $7.34 \pm 0.03$ 
 & $7.159 \pm 0.004$ & $7.165 \pm 0.009$ \\
 PE & $2545\pm 2$ & $2545 \pm 3$  
 &  $2516.9 \pm 0.8$ & $2518.0 \pm 0.8$\\
 \hline 
 \end{tabular}
 \end{table}
 
\section{Conclusion}
In this work, it has been shown that is is possible to represent missing parts of a numerical model using a neural network. 
This was done by creating a hybrid model that simulated a shallow-water model.
We showed that the hybrid model, used for a long term simulation, produced outputs with the same physical properties as the reference physical model. The focus of this paper was on long term simulations, but it was shown that the hybrid model also has some short-term predictive skills up to few months.

One important feature to be stressed is the necessity to add some physical constraints to the neural network output (in our case boundary conditions) to ensure the convergence of the neural network model on long-term simulations. The addition of physical constraints is the only point of the methodology that is specific to our shallow-water application. The approach can be seen as a more general framework to produce hybrid models. Our experiments suggest that if we aim at short term simulations, the definition of such constraints may not be necessary.

The possibility of merging a physical based and a data based model depends strongly on the availability of data and uncertainties in our numerical model. In some cases, observations are indirect or incomplete. So it can be necessary to apply inverse methods or interpolation techniques to produce the data used as training by the neural network. These methods have they own inaccuracies, but it has been shown that neural networks were able to learn the unknown process in the presence of noise.

The hybrid model has demonstrated its ability to reproduce long term simulation in conditions similar to what was encountered during the training phase. But more importantly, even if the exterior forcing (in our case, the wind stress) differs from the one used in the training data, the neural network model still reproduces, with no significant difference, the characteristics of the physical system.

\acknowledgments
JB is grateful to M. Bocquet, A. Carrassi and L. Bertino for useful discussions.
This work was funded by the APPLE-DOM project PNTS-2018-2. JB has been funded by the project REDDA (\#250711) of the Norwegian Research Council.

\listofchanges

\end{document}